\documentclass[graybox]{svmult}

\usepackage{mathptmx}       
\usepackage{helvet}         
\usepackage{courier}        
\usepackage{type1cm}        
\usepackage{makeidx}         
\usepackage{graphicx}        
\usepackage{multicol}        
\usepackage[bottom]{footmisc}

\def\arcs{\hbox{$^{\prime\prime}\,$}}
\def\farcs{\hbox{$^{\prime\prime}\!\!\!.\,$}}

\makeindex             

\begin{document}

\title*{The SKA and ``High-Resolution'' Science}
\titlerunning{The SKA and ``High-Resolution'' Science}
\author{A.P. Lobanov}
\authorrunning{Lobanov} 
\institute{Max-Planck-Institut f\"ur Radioastronomie, Auf dem H\"ugel 69, 53121 Bonn, Germany. \email{alobanov@mpifr.de}}

\maketitle

\abstract{``High-resolution'', or ``long-baseline'', science with the
SKA and its precursors covers a broad range of topics in
astrophysics. In several research areas, the coupling between improved
brightness sensitivity of the SKA and a sub-arcsecond resolution would
uncover truly unique avenues and opportunities for studying extreme
states of matter, vicinity of compact relativistic objects, and
complex processes in astrophysical plasmas. At the same time, long
baselines would secure excellent positional and astrometric
measurements with the SKA and critically enhance SKA image fidelity at
all scales. The latter aspect may also have a substantial impact on
the survey speed of the SKA, thus affecting several key science
projects of the instrument.}

\section{Introduction}
\label{sec:1}

The benchmark design for the SKA Phase 1 \cite{schilizzi2007},
envisaging operations in the 0.3-10\,GHz range and on baselines of
up to several hundred kilometres, would have enabled addressing a
range of scientific areas relying on sub-arcsecond resolution,
including astrometry, pulsar proper motions, supernovae, astrophysical
masers, nuclear regions of AGN, physics of relativistic and mildly
relativistic outflows, kinetic feedback from AGN, evolution of
supermassive black holes and their host galaxies
\cite{carilli2004}. The revised specifications for the SKA$_1$
\cite{dewdney2010,garrett2010}, shifting the operational frequency
range to 0.07-3\,GHz and limiting the baseline length to 100\,km,
leads to a reduction of the instrumental resolution to
0{\farcs}3--1{\farcs}4 for the dishes (in the 0.45--3\,GHz range) and
1{\arcs}--8{\arcs} for the aperture array (in the 0.07--0.45\,GHz
range). 

The ``stand alone'' resolution of SKA$_1$ will therefore be sufficient
for addressing only a subset of topics listed above. Achieving a higher
resolution would rely on inclusion of external antennas and operating
in the VLBI mode. This would be mostly feasible for the dish part of
SKA$_1$, as most of the present day VLBI arrays are operating at
frequencies above 600\,MHz, and there are no definite plans to extend
VLBI operations to below 300\,MHz.  SKA$_1$ operating in Australia can
be an integral part of the LBA/NZ Network and EAVN. It would also have a
somewhat limited common visibility with the VLBA.  SKA$_1$ sited in
South Africa will be a natural partner to EVN+ antennas. In both
cases, collaboration with geodetic VLBI is possible, if 2.3\,GHz will
be maintained as a network frequency by the IVS. 

In addition, the sub-arcsecond resolution of SKA$_1$ may actually be
an essential requirement also for achieving the specifications
envisaged for the traditional ``low-resolution'' science, including
the surveying capabilities of the array (often viewed as a backbone of
the instrument). 

These two aspects of the relevance of long baselines to achieving the
scientific goals of SKA$_1$ are discussed below, with
Section~\ref{apl-sec2} describing potential areas of broad scientific
impact of high-resolution studies with SKA$_1$ and SKA Precursors and
Section~\ref{apl-sec3} discussing the effect of long baselines on
the quality of imaging and surveying capability of SKA$_1$.

\section{High-resolution science with SKA and SKA Precursors}
\label{apl-sec2}

With its present design specification, SKA$_1$ will be most effective
addressing the following subset of topics mentioned above: studies of
galactic and extragalactic supernova remnants, detection of atomic and
molecular gas in galaxies, localisation of non-thermal continuum
production sites in AGN, investigations of extragalactic outflows and
their role in AGN feedback, studies of supermassive black holes and
their relation to galaxy evolution, and research in AGN relic
activity. 

\subsection{Supernova science}

High-sensitivity radio observations at a sub-arcsecond resolution
provide an effective tool to detect and monitor extragalactic SN/SNR
\cite{beswick2006,lonsdale2006,mcdonald2001}, giving a good estimate
of star-formation rate in target galaxies and helping assess the
connection between AGN and star formation.  These measurements rely on
highly sensitive long baselines, and the improvements in sensitivity
already enable detecting and imaging much weaker supernova remnants in
the Milky Way as well \cite{bietenholz2010,mezcua2010}. SKA$_1$ will
enable detecting weaker SN/SNR in a much wider range of galaxies.
The combination of high resolution and superb brightness sensitivity would
enable much longer tracing of evolving supernova shells and supernova
remnants, yielding essential information about their ages and galactic
environment.

\subsection{Atomic and molecular gas in galaxies}

In addition to observations of line emission from H{\,\scriptsize I}
(at 1.42\,GHz) and D{\,\scriptsize I} (at 0.327\,GHz), studies of OH
megamasers (at 1.67\,GHz) and H{\,\scriptsize I}/OH absorption made
with SKA$_1$ will provide a wealth of information about atomic and
molecular gas in the nuclear regions of galaxies
\cite{carilli2004,lobanov2005}.

Extragalactic OH megamasers are detected towards IR luminous galaxies
and they are 10$^3$--10$^6$ times stronger than brightest Galactic
masers \cite{kloeckner2005}, and they are reported to have a
two-component structure \cite{kloeckner2003} possibly tracing an
interaction between the ionisation cones of the nuclear outflow and
the molecular torus. Combining SKA$_1$ with antennas on 3000+\,km
baselines would broaden spectacularly the scope of OH megamaser
studies.

Absorption due to several species, most notably H{\,\scriptsize{I}}
and OH toward compact continuum sources is a unique tool to probe
nuclear regions on parsec scales -- still beating the resolution and
accuracy of optical integral field spectroscopy studies
\cite{mundell2003,peck2001}.  In extragalactic objects, OH absorption
has been used to probe the conditions in warm neutral gas
\cite{kloeckner2005}, and CO and H{\,\scriptsize{I}} absorption has
become a tool of choice to study the molecular tori \cite{pedlar2004}
and interactions between outflows and ambient ISM
\cite{morganti2009,peck2001}.  Studies of the nuclear absorption will
benefit enormously from highly sensitive baselines provided by using
SKA$_1$ alone or together with VLBI arrays.

\subsection{Localisation of non-thermal continuum in AGN}

Detailed knowledge of the mechanism of high-energy particle and
emission production in AGN is pivotal for studies of galactic activity
\cite{lobanov2010} as well as for understanding the kinetic and
radiation feedback from AGN influencing cosmological galaxy
evolution, black hole growth, and large-scale structure formation.

Self-consistent physical models for non-thermal continuum production
in AGN require accurate information about the sites where the bulk of
the high-energy emission is produced. Present arguments place these
sites anywhere between $\sim$1000 gravitational radii of the central
black hole \cite{acciari2009} and a $\sim$100\,pc separation from it
\cite{cheung2007}

Joint modelling of radio, optical, and X-ray data in 3C120 indicates
that radio flares and X-ray dips seem to originate near the accretion
disk \cite{chatterjee2009,marscher2008}, while optical and high-energy
flares are produced in stationary shocks located at $\sim$1 pc
downstream in the jet \cite{arshakian2010,leon2010,schinzel2010}. One
important conclusion from this work is that instantaneous SEDs are
likely to result from several physically different plasma components,
which may lead to considerable difficulties in their interpretation
being made without any reference to spatial location of the emission
observed in different bands.

The combination of broad-band monitoring and high-resolution radio
observations remains the only viable tool for spatial localisation of
non-thermal continuum in AGN. Substantial improvements of sensitivity
and time coverage of such combined programs can be achieved with
SKA$_1$ working as part of VLBI experiments, and it would be certainly
provide essential information for understanding the physics of
high-energy emission production in AGN.

\subsection{Outflows and feedback in AGN}

Evidence is abound for feedback from AGN to play an important role in
physical processes at intergalactic and intracluster scales
\cite{binney1995,nipoti2005}, with the efficiency and mechanism of the
kinetic feedback from nuclear outflows still being poorly understood.

SKA$_1$ will be an excellent tool for probing physical conditions in
low-energy tail of outflowing plasma which is believed to carry the
bulk of kinetic power of the outflow. This will enable making detailed
quantitative studies of evolution and re-acceleration of non-thermal
plasma in cosmic objects and provide essential clues for understanding
the power and efficiency of the kinetic feedback from AGN and its
effect on activity cycles in galaxies and cosmological growth of
supermassive black holes.  Such studies are critically needed for
making a detailed assessment of the role played by AGN in the
formation and evolution of the large-scale structure in the Universe.

\subsection{Galactic mergers and supermassive black holes}

High-resolution and high-sensitivity radio observations are expected
to provide arguably the best AGN and SMBH census up to very high
redshifts \cite{carilli2004}. This will enable cosmological studies of
SMBH growth, galaxy evolution, and the role played by galactic mergers
in nuclear activity and SMBH evolution.

Most powerful AGN are produced by galactic/SMBH mergers
\cite{dimatteo2004,haehnelt2002}. Activity is reduced when a loss cone
is formed and most of nuclear gas is accreted onto SMBH
\cite{merritt2005}. The remaining secondary SMBH helps maintaining
activity of the primary \cite{dokuchaev1991}, and the evolution of
nuclear activity can be connected to the dynamic evolution of binary
SMBH in galactic centres \cite{lobanov2008}. Direct detections of secondary
SMBH in post-merger galaxies are the best way to the evolution of
black holes and galaxies together. Some of the secondary BH may be
``disguised'' as ULX objects \cite{mezcua2010} accreting at ~10-5 of
the Eddington rate.  They are not detected in deep radio images at
present.

SKA$_1$ would be a superb tool for detecting and classifying such
objects, thus providing an essential observational information about
the SMBH evolution in post-merger galaxies and its influence on the
galactic activity, formation of collimated outflows and feedback from
AGN.

\subsection{Radio relics and AGN cycles}

Nuclear activity in galaxies is believed to be episodic or
intermittent, with estimates of activity cycles reaching up to $10^8$
years~\cite{komissarov1994,stanghellini2005}. The episodes of activity
are related closely to mergers of galaxies~\cite{schoenmakers2000} and
evolution of supermassive binary black holes resulting from galactic
mergers~\cite{lobanov2008,merritt2005}. Both the onset and the latest
stages of the jet activity are poorly studied at the moment, because
in either case the flowing plasma emit largely at low frequencies.

Relics of previous cycles of nuclear activity are difficult to detect
at centimetre wavelengths because of significant losses due to
expansion and synchrotron emission. At centimetre wavelengths, such
relics decay below the sensitivity limits of the present-day
facilities within $10^4$--$10^5$ years after the fuelling of extended
lobes stops. This explains the relatively small number of such relics
known so far. SKA$_1$, working below 1\,GHz, would be able to detect
such relics for at least $10^7$ years after the fuelling stops, and
this would make it possible to assess the activity cycles in a large
number of objects, searching for signs of re-started activity in
radio-loud objects and investigating ``paleo'' activity in presently
radio-quiet objects. This information will be essential for
constructing much more detailed models of evolution and nuclear
activity of galaxies.

\section{Long baselines for ``low resolution'' science}
\label{apl-sec3}

Availability of long baselines is not only a definitive requirement
for a number of science areas relying on high-resolution imaging, but
also an important factor for achieving design specifications envisaged
for several key science areas of SKA$_1$ and full SKA. This concerns
primarily the design goals for the survey speed and r.m.s. sensitivity
to extended, low-surface emission.

\subsection{Imaging with the SKA}

In the SKA developments, imaging capability is often viewed as a
``tradeoff'' against the survey speed \cite{cordes2009} -- hence
``core-spread'' array configurations are favoured and long baselines
are downgraded \cite{dewdney2010,garrett2010}.  But the two-order of
magnitude sensitivity improvement envisaged for SKA will lead to
situation when surveying is made in crowded fields with a number of
resolved objects per primary beam of the receiving element.

In this circumstance, high fidelity imaging becomes then an essential
feature rather than a ``tradeoff'', and the combination
$A_\mathrm{eff}/T_\mathrm{sys}$ becomes inadequate as a figure of
merit describing the r.m.s. sensitivity and survey speed,
$S_\mathrm{s} \propto (A_\mathrm{eff}/T_\mathrm{sys})^2$.

It has therefore been argued \cite{bunton2006,lobanov2003} that SKA
needs to have the capability of imaging adequately (at least) all
those spatial frequencies at which there is more than one sky object
per primary beam.

\subsection{Survey speed vs. imaging fidelity}

Optimisation for the survey speed as expressed by the
$A_\mathrm{eff}/T_\mathrm{sys}$ factor is based on two implicit
assumptions:
\begin{itemize}
\item[{\em 1)}]~The number of objects $N_\mathrm{source}$ detected in the primary
beam area is small, $N_\mathrm{source} \le (B_\mathrm{max}/D_\mathrm{ant})^2$ (where $D_\mathrm{ant}$ is the diameter of antenna and $B_\mathrm{max}$ is the maximum baseline length);
\item[{\em 2)}]~All these objects are essentially unresolved by the
array, implying that they have angular sizes $\theta_\mathrm{source} \le
2\, \mathrm{FWHM}/\sqrt{\mathrm{SNR}}$.
\end{itemize}
Neither of these two conditions will be realised in the case the case
of both SKA and SKA$_1$. The primary beam area of a 12-m dish will
contain $\sim$50 objects ($S_\mathrm{1.4GHz}>0.4$\,mJy) larger than
$\theta_\mathrm{source}$.  This leads to a strong requirement of
optimisation for dynamic range and high-fidelity imaging (hence the distribution of the collecting area).

This effect  has been investigated in a number of studies
\cite{bunton2006,lal2009,lobanov2003} showing that a uniform
structural sensitivity is desired to alleviate the problems posed by
resolved objects in the field of view. The structural sensitivity can
described by the {\em uv}-gap parameter \cite{lobanov2003}, $\Delta
u/u$. For an array uniformly sensitive to all detectable angular
scales, $\Delta u/u = const$ on all baselines.
The structural sensitivity of an array can be described by a factor
\[
\eta_{uv} = \exp \left[ \frac{\pi^2}{16\ln\, 2}\frac{\Delta u}{u} 
\left( \frac{\Delta u}{u} +2 \right) \right]^{-1}\, ,
\]
with $\eta_{uv} \equiv 1$ for a filled aperture (for which $\Delta
u/u \equiv 0$). An imperfect {\em uv}-coverage of an interferometric
array results in an additional, scale-dependent factor of the image
noise, $\sigma_{uv} = \sigma_\mathrm{rms} / \eta_{uv}$, and this has
to be taken into account when estimating the performance of SKA for
surveys. These estimates should be done using the factor $\eta_{uv} (A_\mathrm{eff}/T_\mathrm{sys})$.

\subsection{Configuration choices}

Analysis based on the structural sensitivity factor shows that,
compared to the ``core-spread'' array configuration (adopted as a
benchmark for SKA$_1$ and SKA, a logarithmic
\cite{bunton2006,lobanov2003} or gaussian array of the same extent in
baseline length will have an approximately 3 times worse brightness
sensitivity on scales of $\le 0.1\,B_\mathrm{max}$, a $\sim 30\%$
lower noise in snapshot images, and ~a $\sim 2$ times lower sidelobe
level. For a 10 times larger gaussian/logarithmic array, these figures
are reduced only slightly, while the confusion limit is lowered by a
factor of 100.  These arguments show that adopting such a
configuration is rather a necessity for large area surveys. In
contrast, for the core-spread configuration, the noise increase due to
$\sigma_{uv}$ would have to be offset by increasing the observing by a
factor of 10(!) if one would require to reach within 5\% of the
r.m.s. specification based on the simple
$A_\mathrm{eff}/T_\mathrm{sys}$ factor.

\section{Conclusions}

SKA$_1$ would be able to address a number of important astrophysical
areas of study relying on high-resolution radio observations: studying
supernovae; providing a good account of starburst activity in
galaxies; using megamasers and nuclear absorption to probe the nuclear
gas in galaxies; understanding in detail the physics of (ultra- and
mildly-relativistic) outflows and their connection to the nuclear
regions in galaxies; searching for radio emission from weaker AGN and
secondary black holes in post-merger galaxies; and addressing the
questions of relic activity and activity cycles in AGN.  Reliable high
resolution imaging is also needed for achieving the scientific goals
in the key science areas of SKA$_1$ requiring low-resolution imaging
and surveying of large areas in the sky, most notably the science
areas considered as the very ``selling point'' of the instrument.

\bibliographystyle{spmpsci}
\bibliography{lobanov}

\begin{thebibliography}{10}
\providecommand{\url}[1]{{#1}}
\providecommand{\urlprefix}{URL }
\expandafter\ifx\csname urlstyle\endcsname\relax
  \providecommand{\doi}[1]{DOI~\discretionary{}{}{}#1}\else
  \providecommand{\doi}{DOI~\discretionary{}{}{}\begingroup
  \urlstyle{rm}\Url}\fi

\bibitem{acciari2009}
{Acciari}, V.A., {Aliu}, E., {Arlen}, T., {Bautista}, M., {Beilicke}, M.,
  {Benbow}, W., {Bradbury}, S.M., {Buckley}, J.H., {Bugaev}, V., {Butt}, Y.,
  et~al.: {Radio Imaging of the Very-High-Energy {$\gamma$}-Ray Emission Region
  in the Central Engine of a Radio Galaxy}.
\newblock Science \textbf{325}, 444 (2009)

\bibitem{arshakian2010}
{Arshakian}, T.G., {Le{\'o}n-Tavares}, J., {Lobanov}, A.P., {Chavushyan}, V.H.,
  {Shapovalova}, A.I., {Burenkov}, A.N., {Zensus}, J.A.: {Observational
  evidence for the link between the variable optical continuum and the
  subparsec-scale jet of the radio galaxy 3C 390.3}.
\newblock MNRAS \textbf{401}, 1231 (2010)

\bibitem{beswick2006}
{Beswick}, R.J., {Riley}, J.D., {Marti-Vidal}, I., {Pedlar}, A., {Muxlow},
  T.W.B., {McDonald}, A.R., {Wills}, K.A., {Fenech}, D., {Argo}, M.K.: {15
  years of very long baseline interferometry observations of two compact radio
  sources in Messier 82}.
\newblock MNRAS \textbf{369}, 1221 (2006)

\bibitem{bietenholz2010}
{Bietenholz}, M.F., {Bartel}, N., {Milisavljevic}, D., {Fesen}, R.A.,
  {Challis}, P., {Kirshner}, R.P.: {The first VLBI image of the young,
  oxygen-rich supernova remnant in NGC 4449}.
\newblock MNRAS \textbf{409}, 1594 (2010)

\bibitem{binney1995}
{Binney}, J., {Tabor}, G.: {Evolving Cooling Flows}.
\newblock MNRAS \textbf{276}, 663 (1995)

\bibitem{bunton2006}
{Bunton}, J.: {Area Scaling for the SKA}.
\newblock {SKA Memo Series}~79, SPDO (2006)

\bibitem{carilli2004}
{Carilli}, C., {Rawlings}, S. (eds.): {Science with the Square Kilometre
  Array}. {Elsevier} (2004)

\bibitem{chatterjee2009}
{Chatterjee}, R., {Marscher}, A.P., {Jorstad}, S.G., {Olmstead}, A.R.,
  {McHardy}, I.M., {Aller}, M.F., {Aller}, H.D., et~al.: {Disk-Jet Connection
  in the Radio Galaxy 3C 120}.
\newblock ApJ \textbf{704}, 1689 (2009)

\bibitem{cheung2007}
{Cheung}, C.C., {Harris}, D.E., {Stawarz}, {\L}.: {Superluminal Radio Features
  in the M87 Jet and the Site of Flaring TeV Gamma-Ray Emission}.
\newblock ApJL \textbf{663}, L65 (2007)

\bibitem{cordes2009}
{Cordes}, J.M.: {Survey Metrics}.
\newblock {SKA Memo Series} 109, SPDO (2009)

\bibitem{dewdney2010}
{Dewdney}, P., {bij de Vaate}, J.G., {Cloete}, K., {Gunst}, A., {Hall}, D.,
  {McCool}, R., {Roddis}, N., {Turner}, W.: {SKA Phase 1: Preliminary System
  Description}.
\newblock {SKA Memo Series} 130, SPDO (2010)

\bibitem{dimatteo2004}
{Di Matteo}, T., {Croft}, R.A.C., {Springel}, V., {Hernquist}, L.: {The
  Cosmological Evolution of Metal Enrichment in Quasar Host Galaxies}.
\newblock ApJ \textbf{610}, 80 (2004)

\bibitem{dokuchaev1991}
{Dokuchaev}, V.I.: {Joint evolution of a galactic nucleus and central massive
  black hole}.
\newblock MNRAS \textbf{251}, 564 (1991)

\bibitem{garrett2010}
{Garrett}, M.A., {Cordes}, J.M., {De Boer}, D., {Jonas}, J.L., {Rawlings}, S.,
  {Schilizzi}, R.T.: {Concept Design for SKA Phase 1 (SKA$_1$)}.
\newblock {SKA Memo Series} 125, SPDO (2010)

\bibitem{haehnelt2002}
{Haehnelt}, M.G., {Kauffmann}, G.: {Multiple supermassive black holes in
  galactic bulges}.
\newblock MNRAS \textbf{336}, L61 (2002).
\newblock \doi{10.1046/j.1365-8711.2002.06056.x}

\bibitem{kloeckner2005}
{Kl{\"o}ckner}, H., {Baan}, W.A.: {Understanding Extragalactic Hydroxyl}.
\newblock Astrop. \& Space Sci. \textbf{295}, 277 (2005)

\bibitem{kloeckner2003}
{Kl{\"o}ckner}, H., {Baan}, W.A., {Garrett}, M.A.: {Investigation of the
  obscuring circumnuclear torus in the active galaxy Mrk231}.
\newblock Nature \textbf{421}, 821 (2003)

\bibitem{komissarov1994}
{Komissarov}, S.S., {Gubanov}, A.G.: {Relic radio galaxies: evolution of
  synchrotron spectrum}.
\newblock A\&A \textbf{285}, 27 (1994)

\bibitem{lal2009}
{Lal}, D.V., {Lobanov}, A.P., {Jim\'enez-Monferrer}, S.: {Array configuration
  studies for the Square Kilometre Array -- Implementation of figures of merit
  based on spatial dynamic range}.
\newblock {SKA Memo Series} 107, SPDO (2009)

\bibitem{leon2010}
{Le{\'o}n-Tavares}, J., {Lobanov}, A.P., {Chavushyan}, V.H., {Arshakian}, T.G.,
  {Doroshenko}, V.T., {Sergeev}, S.G., {Efimov}, Y.S., {Nazarov}, S.V.:
  {Relativistic Plasma as the Dominant Source of the Optical Continuum Emission
  in the Broad-Line Radio Galaxy 3C 120}.
\newblock ApJ \textbf{715}, 355 (2010)

\bibitem{lobanov2003}
{Lobanov}, A.P.: {Imaging with the SKA: Comparison to other future major
  instruments}.
\newblock {SKA Memo Series}~38, SPDO (2003)

\bibitem{lobanov2005}
{Lobanov}, A.P.: {Radio spectroscopy of active galactic nuclei}.
\newblock MemSAIS \textbf{7}, 12 (2005)

\bibitem{lobanov2008}
{Lobanov}, A.P.: {Binary supermassive black holes driving the nuclear activity
  in galaxies .}
\newblock MemSAI \textbf{79}, 1306 (2008)

\bibitem{lobanov2010}
{Lobanov}, A.P.: {Physical properties of blazar jets from VLBI observations}.
\newblock ArXiv:1010.2856  (2010)

\bibitem{lonsdale2006}
{Lonsdale}, C.J., {Diamond}, P.J., {Thrall}, H., {Smith}, H.E., {Lonsdale},
  C.J.: {VLBI Images of 49 Radio Supernovae in Arp 220}.
\newblock ApJ \textbf{647}, 185 (2006)

\bibitem{marscher2008}
{Marscher}, A.P., {Jorstad}, S.G., {D'Arcangelo}, F.D., {Smith}, P.S.,
  {Williams}, G.G., {Larionov}, V.M., et~al.: {The inner jet of an active
  galactic nucleus as revealed by a radio-to-{$\gamma$}-ray outburst}.
\newblock Nature \textbf{452}, 966 (2008)

\bibitem{mcdonald2001}
{McDonald}, A.R., {Muxlow}, T.W.B., {Pedlar}, A., {Garrett}, M.A., {Wills},
  K.A., {Garrington}, S.T., {Diamond}, P.J., {Wilkinson}, P.N.: {Global very
  long-baseline interferometry observations of compact radio sources in M82}.
\newblock MNRAS \textbf{322}, 100 (2001)

\bibitem{merritt2005}
{Merritt}, D., {Milosavljevi{\'c}}, M.: {Massive Black Hole Binary Evolution}.
\newblock Living Reviews in Relativity \textbf{8}, 8 (2005)

\bibitem{mezcua2010}
{Mezcua}, M., {Lobanov}, A.P.: {Compact radio emission in Ultraluminous X-ray
  sources}.
\newblock ArXiv:1011.0946  (2010)

\bibitem{morganti2009}
{Morganti}, R., {Peck}, A.B., {Oosterloo}, T.A., {van Moorsel}, G., {Capetti},
  A., {Fanti}, R., {Parma}, P., {de Ruiter}, H.R.: {Is cold gas fuelling the
  radio galaxy NGC 315?}
\newblock A\&A \textbf{505}, 559 (2009)

\bibitem{mundell2003}
{Mundell}, C.G., {Wrobel}, J.M., {Pedlar}, A., {Gallimore}, J.F.: {The Nuclear
  Regions of the Seyfert Galaxy NGC 4151: Parsec-Scale H I Absorption and a
  Remarkable Radio Jet}.
\newblock ApJ \textbf{583}, 192 (2003)

\bibitem{nipoti2005}
{Nipoti}, C., {Binney}, J.: {Time variability of active galactic nuclei and
  heating of cooling flows}.
\newblock MNRAS \textbf{361}, 428 (2005)

\bibitem{peck2001}
{Peck}, A.B., {Taylor}, G.B.: {Evidence for a Circumnuclear Disk in 1946+708}.
\newblock ApJL \textbf{554}, L147 (2001)

\bibitem{pedlar2004}
{Pedlar}, A., {Muxlow}, T., {Smith}, R., {Thrall}, H., {Beswick}, R., {Aalto},
  S., {Booth}, R., {Wills}, K.: {OH Molecules and Masers in Messier 82}.
\newblock In: {S.~Aalto, S.~Huttemeister, \& A.~Pedlar} (ed.) The Neutral ISM
  in Starburst Galaxies, \emph{Astronomical Society of the Pacific Conference
  Series}, vol. 320, p. 183 (2004)

\bibitem{schilizzi2007}
{Schilizzi}, R.T., {Alexander}, P., {Cordes}, J.M., {Dewdney}, P.E., {Ekers},
  R.D., {Gaensler}, B.M., {Faulkner}, A.J., {Hall}, P.J., {Jonas}, J.L.,
  {Kellermann}, K.I.: {Preliminary Specifications for the Square Kilometre
  Array}.
\newblock {SKA Memo Series} 100, SPDO (2007)

\bibitem{schinzel2010}
{Schinzel}, F.K., {Lobanov}, A.P., {Jorstad}, S.G., {Marscher}, A.P., {Taylor},
  G.B., {Zensus}, J.A.: {Radio Flaring Activity of 3C 345 and its Connection to
  Gamma-ray Emission}.
\newblock ArXiv:1012.2820  (2010)

\bibitem{schoenmakers2000}
{Schoenmakers}, A.P., {de Bruyn}, A.G., {R{\"o}ttgering}, H.J.A., {van der
  Laan}, H., {Kaiser}, C.R.: {Radio galaxies with a `double-double morphology'
  - I. Analysis of the radio properties and evidence for interrupted activity
  in active galactic nuclei}.
\newblock MNRAS \textbf{315}, 371 (2000)

\bibitem{stanghellini2005}
{Stanghellini}, C., {O'Dea}, C.P., {Dallacasa}, D., {Cassaro}, P., {Baum},
  S.A., {Fanti}, R., {Fanti}, C.: {Extended emission around GPS radio sources}.
\newblock A\&A \textbf{443}, 891 (2005)

\end{thebibliography}

\end{document}